\documentclass[twocolumn, showpacs,apl,amsmath,amssymb,amsthm,superscriptaddress]{revtex4}

\usepackage{graphicx,epsf,epsfig,ulem}
\usepackage{epsfig}
\usepackage{bm}
\usepackage{bbm}
\usepackage{subfigure}
\usepackage{color}

\begin{document}

\title{SiGe/Si quantum dot electron spin decoherence dependence on $^{73}$Ge}

\author{Wayne M. \surname{Witzel} \email{wwitzel@sandia.gov}}
\affiliation{Sandia National Laboratories, New Mexico 87185 USA}
\author{Rajib \surname{Rahman}}
\affiliation{Sandia National Laboratories, New Mexico 87185 USA}
\author{Malcolm S. \surname{Carroll}}
\affiliation{Sandia National Laboratories, New Mexico 87185 USA}

\begin{abstract}
We theoretically study the nuclear spin induced decoherence of a quantum dot in Si that is confined at a SiGe interface.  We calculate decoherence time dependence on $^{73}$Ge in the barrier layer to evaluate the importance of Ge as well as Si enrichment for long decoherence times.
We use atomistic tight-binding modeling for an accurate account of the electron wavefunction which is particularly important for determining the contact hyperfine interactions with the Ge nuclear spins.  We find decoherence times due to Ge spins at natural concentrations to be milliseconds.  This suggests SiGe/Si quantum dot devices employing enriched Si will require enriched Ge as well in order to benefit from long coherence times.  We provide a comparison of $T_2$ times for various fractions of nonzero spin isotopes of Si and Ge.
\end{abstract}

\maketitle

\section{Introduction}
An electron spin in silicon is a promising candidate for a quantum
bit.  Sources of decoherence such as spectral diffusion of nuclear
spins can be nearly eliminated through isotopic enrichment and other
decoherence mechanisms such as spin-orbit coupling are small.  Quantum dots offer a
way to control and couple spins electrically, demonstrated in GaAs,
and have shown fast single qubit gate operations.  Significant
progress towards achieving coherent control of spins in single and
double quantum dots in Si/SiGe have been suggested
recently~\cite{NakulShaji2008, Maune2011, ZhanShi2011, ThalakulamSiGeQdot, BorselliSiGeDQD}.  Long
decoherence times in silicon have been achieved~\cite{Tyryshkin50ppm}, in large part,
through enrichment of the Si with the zero nuclear spin isotope $^{28}$Si.
In SiGe/Si quantum dots, the wavefunction is in close proximity with
a distribution of germanium nuclei in the SiGe barrier region.
Approximately 7.8 percent of the isotopes in naturally occurring pure
germanium carry spin (i.e., $^{73}$Ge).  In this article we use a
tight-binding method, NEMO3D~\cite{Klimeck1}, to calculate the spatial extent of
the Si/SiGe quantum dot wavefunction and spin-bath cluster methods to
calculate the decoherence behavior resulting from the presence of
these non-zero, 9/2, nuclear spins.

\section{Electron wavefunction calculation}
The spatial extent of an electron localized in a quantum dot is calculated using an atomistic tight binding (TB) approach embedded in the Nanoelectronic Modeling Tool (NEMO3D) \cite{Klimeck1}, which solves the single electron eigen spectra for multi-million atom systems. We adopt this detailed atomistic approach to model a realistic quantum dot as the TB method in general captures a variety of atomic scale effects. In particular, this quantitative description is needed to accurately obtain the wavefunction amplitudes on the Ge atoms, responsible for the contact hyperfine interaction \cite{Park}. A good description of the spatial distribution of the wavefunction is also necessary to compute the dipolar interactions precisely \cite{Park}.     

We simulate a quantum dot confined vertically in a SiGe/Si quantum well and laterally by a harmonic potential, as described by the electronic (el) Hamiltonian,
\begin{equation}
 H^{\mbox{\scriptsize el}} = H^{\mbox{\scriptsize el}}_0+a(x-x_0)^2+a(y-y_0)^2+e F_z z
 \label{eq1}
\end{equation} 
\noindent
where $H_0$ is the TB Hamiltonian of the Si-SiGe lattice. The 2nd and 3rd terms represent a 2D parabolic quantum dot of curvature $a=3.137 \times 10^{-7}$ eV/nm$^2$, which translates to a characteristic ground state energy $E=0.5$ meV and a dot radius of $R=25$ nm (i.e. the distance at which the wavefunction amplitude falls off as $1/e$ of the peak value). The last term in Eq.~(\ref{eq1}) is due to a variable vertical E-field $F_z$, which simulates the effect of a top enhancement gate \cite{Lu}. A harmonic potential represents a reasonably good estimate of the confinement for electrostatically gated structures \cite{Nielsen}. T
While minor numerical variations of the results presented here are expected for much smaller dot sizes, in which the local alloy disorder has greater impact from sample to sample, the same trends and order of magnitude estimates will remain valid. 

The Hamiltonian of the host materials $H_0$ is formulated semi-empirically with the 10 band sp$^3$d$^5$s* nearest neighbor model \cite{Klimeck1} to represent the full bandstructure of the hetero-structure. The TB model parameters are well-established in literature \cite{Boykin}, and validated in a variety of earlier works \cite{Klimeck2, Park}. The model of the SiGe region is purely atomistic \cite{Boykin}, and no virtual crystal approximations are involved.    

The vertical structure is a 10 nm thick, tensile strained silicon layer sandwiched between fully relaxed SiGe barrier layers of thickness 41 nm each, above and below the strained Si layer respectively, as shown in Fig. 1(a). The equilibrium lattice constants and strain in the two layers are also calculated through minimization of the strain energy calculated using the valence-force field Keating model \cite{Keating}. The size of the lattice is chosen so that the SiGe, to a good approximation, is fully relaxed \cite{Kharche}. For strain relaxation, simulations of 22 million atoms were performed, whereas electronic structure simulations could be done on a reduced domain of 3.5 million atoms incorporating only a part of the SiGe regions near the strained Si region.
The band offsets between Si$_{1-x}$Ge$_x$ and Si obtained from atomistic tight-binding simulations compare well with experimentally known values~\cite{Jiang}.

The full TB Hamiltonian, including the electrostatic potential given in Eq.~(\ref{eq1}), is solved by a parallel Lanczos algorithm to obtain a number of the lowest energy states of the quantum dot. The ground state wavefunction of the QD is of relevance in this work.  Since Si has a sixfold degenerate conduction band, the orbital states are multiplied by six in general. However, presence of strain, vertical E-field, and the SiGe/Si interface break the conduction band valley degeneracy of the system~\cite{BoykinValley, FriesenValley}, introducing an energy gap between the ground state and the excited states.  The splitting between the lowest two valleys increases with the vertical E-field~\cite{BoykinValley, SaraivaValley}.  Our consideration in this Article is for a spin qubit encoding and we assume that the excited valley states are kept much further than $k_B T$ above the two energy levels of the qubit states.  We calculate valley splitting in our model to range from $0.1$ to $1.2$~meV in correspondence with our E-field range of $1-20$~MV/m. This splitting is ample for double quantum dot qubits encoded in the spin zero subspace of
singlet and triplet states~\cite{Petta}, which is the predominant system of
consideration for silicon quantum dot based quantum computing. 

\begin{figure}[htbp]
\includegraphics[width=3.4in]{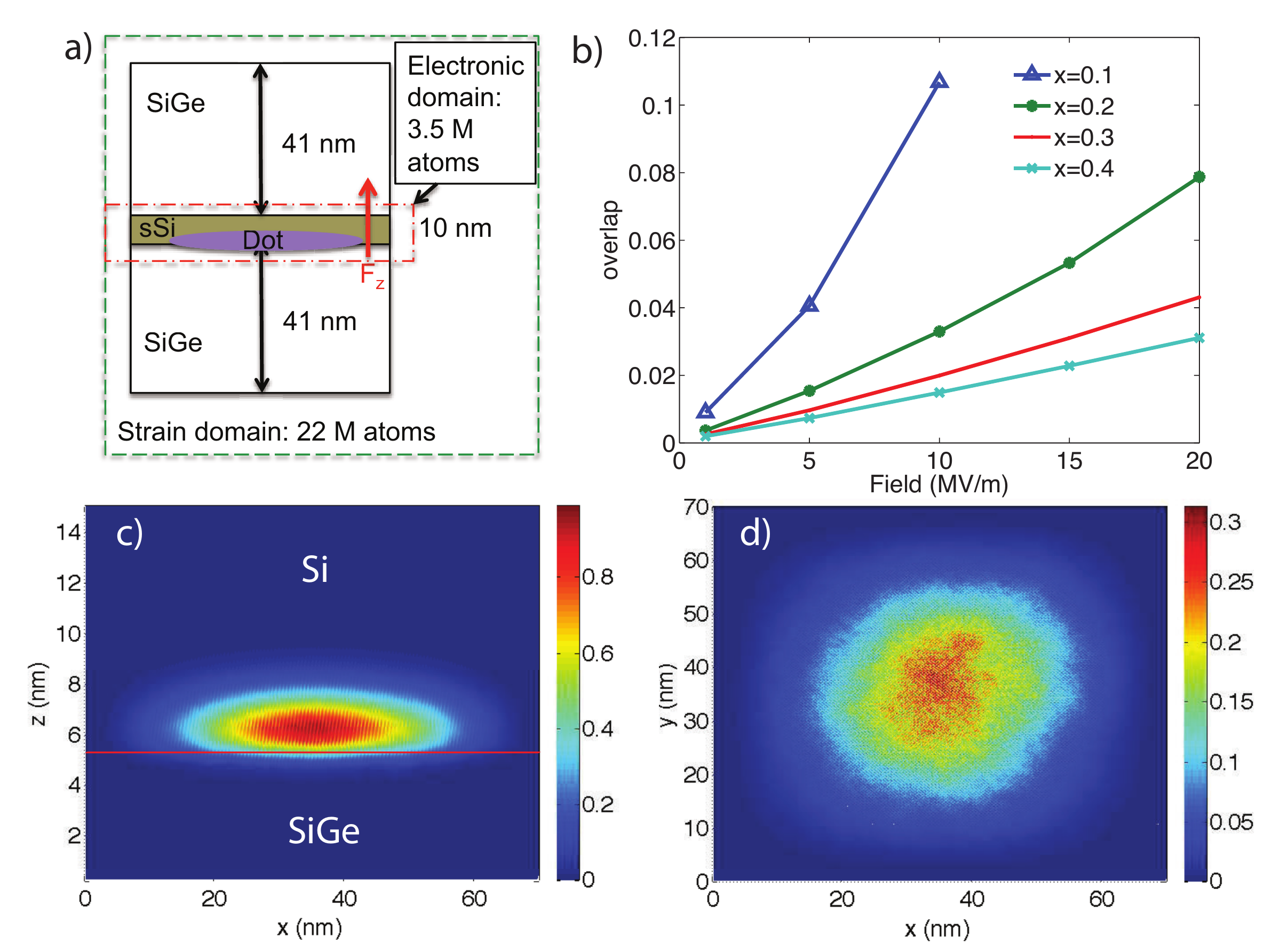}
\caption{
\label{Fig:Wavefunction}
a) The simulated device structure showing the SiGe and Si regions for both strain and electronic structure calculations. b) The expectation value of the wavefunction in the SiGe region as a function of E-field and Ge concentration $x$ in Si$_{1-x}$Ge$_x$. c) Wavefunction probability density plot for the yz plane showing confinement at the interface. d) Wavefunction probability density in the xy plane near the interface inside SiGe showing variations due to alloy disorder. c) and d) are plotted for $F_z=20$ MV/m and a Ge concentration $x=0.3$.
}
\label{fi1}
\end{figure}

The penetration and overlap of the electron wavefunction in the Si$_{1-x}$Ge$_x$ barrier is calculated as a function of both the Ge concentration $x$ in the alloy and the vertical field, as shown in Fig.~\ref{Fig:Wavefunction}~(b). The overlap is defined as the expectation value of the wavefunction in the Si$_{1-x}$Ge$_x$ region. The overlap magnitude changes 5 fold for the entire range of common experimental parameters examined. As expected, the more the Ge concentration, the higher is the interfacial barrier, and less the wavefunction overlap. 
Applying a higher E-field pulls the wavefunction more into the SiGe barrier, and increases the overlap. 
In Fig. 2(c) and 2(d), we show two different cross-sections of the wavefunction, in the xz plane, and in the xy plane near the interface, respectively. The random fluctuations in Fig. 2(d) is due to the alloy disorder in SiGe. The dipole and hyperfine coupling to the nuclear spin bath in the Ge are calculated using the electron wavefunction envelope from the TB solution for each corresponding E-field strength and Ge concentration.

\section{Spin bath cluster expansion calculations}
We examine the central spin decoherence of the quantum dot single electron spin in the silicon quantum well among remotely located $^{73}$Ge nuclear spins in the SiGe barrier.  The quantum dot spin experiences fluctuations of its energy splitting due to spectral diffusion (SD).  To calculate the evolution of the central spin, we use a cluster expansion technique~\cite{WitzelCluster}, which has been shown previously to agree extremely well with decoherence measurements of bulk silicon systems in the presence of varying concentrations of background $^{29}$Si~\cite{Abe10, WitzelDonors}. 

Here we independently consider the effects of various concentrations of either $^{73}$Ge or $^{29}$Si nuclear spins.
We consider isotropic (contact) hyperfine 
and anisotropic hyperfine (AHF) interactions
of the quantum dot electron spin with the nuclei.
We also consider dipolar interactions between the nuclear spins which are responsible
for the dynamics of the nuclear spin bath that causes SD.  
We neglect quadrupole interactions which could be significant for
$^{73}$Ge; these interactions are unknown to our knowledge and its effects
decrease with increasing magnetic field.
Taking the
limit of a large magnetic field such that Zeeman energies
are conserved and only flip-flop dynamics of bath spins are allowed, 
the effective spin (sp) Hamiltonian is
\begin{eqnarray}
H^{\mbox{\scriptsize sp}}_{\mbox{\scriptsize eff}} &=& H^{\mbox{\scriptsize sp}}_{0} + \sum_{n > m} H^{\mbox{\scriptsize sp}}_{n, m}, \\
H^{\mbox{\scriptsize sp}}_0 &=& \sum_{n} A_{n} \hat{S}^z \hat{I}_n^z
- 4 \sum_{n > m} \gamma_n \gamma_m d({\bf R}_n - {\bf R}_m) \hat{I}_n^z \hat{I}_m^z,~~~~ \\
H^{\mbox{\scriptsize sp}}_{n, m} &=& \gamma_n \gamma_m d({\bf R}_n - {\bf R}_m) \left( \hat{I}_n^+ \hat{I}_m^- + \hat{I}_n^- \hat{I}_m^+ \right),~~ \\
\label{Eq:An}
A_n &=& \frac{8 \pi}{3} \gamma_e \gamma_n | \Psi({\bf R}_n)|^2 
- \\ 
\nonumber
&& 4 \gamma_e \gamma_n \sum_{i, {\bf r}_i \neq {\bf R}_n} |\Psi({\bf r}_i)|^2 d({\bf r}_i - {\bf R}_n) \\
d({\bf r}) &=& \frac{1 - 3 (r_z/r)^2}{4 r^3}
\end{eqnarray}
written in atomic units; a factor of $\hbar/(4 \pi \epsilon_0)$ is
implied for the Hamiltonian.
We use $\hat{S}$ for denoted electron spin operators and $\hat{I}_n$ for denoted nuclear spin operators.
For the gyromagnetic ratios, we use $\gamma_e = (g = 2) \mu_B / \hbar
= 1.76 \times 10^{11} \mbox{(T s)$^{-1}$}$, $\gamma_{n} = 0.936 \times
10^{7} \mbox{(T s)$^{-1}$}$ for $^{73}$Ge, and $\gamma_{n} = -5.319
\times 10^{7} \mbox{(T s)$^{-1}$}$ for $^{29}$Si.  The hyperfine
interaction, $A_n$, consists of an isotropic Fermi contact hyperfine
term proportional to the probability of the electron being on top of
the $n$th nucleus and an AHF term that is the dipolar
interaction averaged over the rest of the electron wavefunction. 
This AHF term is negligible compared with
the contact hyperfine for nuclei within the quantum dot region
but dominates far from the quantum dot.
For $| \Psi({\bf R}_n)|^2$ of the contact hyperfine interaction, we
must multiply the envelope of the wavefunction from the TB solution by
a factor $\eta = \| u({\bf R}_n) \|^2$, where $u({\bf R}_n)$ is the
periodic Bloch function.  For the $^{29}$Si, we use $\eta =
186$~\cite{Assali2011}.  For the $^{73}$Ge, we use an $\eta$ that is ten
times larger as a rough estimate~\cite{GeEta}.  The probability
density is higher at $^{73}$Ge sites than at $^{29}$Si since Ge has
more charge.  The $^{29}$Si are spin 1/2 nuclei and the $^{73}$Ge are
spin 9/2 nuclei.

In the large magnetic field limit, the AHF contributions from the 
Bloch part of the wavefunctions have a negligible effect.
Relaxing this limit, these AHF
contributions cause B-field dependent
electron spin echo envelope modulations (ESEEM)~\cite{Reijerse91}
arising from $\hat{S}^{z} \hat{I}^{\pm}$ terms that can be significant
in Si~\cite{HaleAHF}.
These Bloch-dependent AHF interactions strengths are calculated to 
be no more than $3\%$ of the corresponding (per nucleus) contact
hyperfine~\cite{Assali2011} strengths.  
At a B-field of $100~$mT, we estimate
the maximum ESEEM amplitude to be at most $3\%$ in our models with
natural Ge ($0.1\%$ due to natural Si with its smaller $\eta$).
This amplitude maximum scales with nuclear concentration and inversely with
the B-field squared.  
Any other B-field dependence from keeping the
hyperfine and dipolar interaction terms that do not conserve
Zeeman energy are negligible above $10~$mT according to our cluster
calculations.  In our subsequent results, we neglect
B-field dependent effects.  

The Hahn spin echo is used as a standard decoherence measure, $T_2$ commonly reported as the time that the Hahn spin echo reaches $\exp{(-1)}$ times the peak value.  We compute Hahn spin echo decays and associated $T_2$ using a cluster expansion technique~\cite{WitzelCluster}.  The Hahn spin echo consists of initializing the quantum dot electron spin to an equal superposition of up and down, freely evolving for a time $\tau$, applying a spin-flipping $\pi$ rotation, then evolving again for time $\tau$.  We treat the refocusing pulse, the midway $\pi$ rotation, as instantaneous and ideal.  The echo decay is parameterized by the total evolution time $t = 2 \tau$.  The cluster expansion works particularly well where the interactions among bath spins are weak relative to the interactions with the central spin, as is the case since $\gamma_n \ll \gamma_e$.  We are therefore justified in using the lowest order of the cluster expansion, the pair approximation, which accounts for the effects of flip-flopping among all pairs.  This approximation is expressed as
\begin{eqnarray}
\hat{U} &=& e^{-i H^{\mbox{\scriptsize sp}}_{\mbox{\scriptsize eff}} \tau} \hat{S}_x e^{-i H^{\mbox{\scriptsize sp}}_{\mbox{\scriptsize eff}} \tau}, \\
\hat{U}_{n, m} &=& e^{-i (H^{\mbox{\scriptsize sp}}_0 + H^{\mbox{\scriptsize sp}}_{n, m}) \tau} \hat{S}_x e^{-i (H^{\mbox{\scriptsize sp}}_0 + H^{\mbox{\scriptsize sp}}_{n, m}) \tau}, \\
v_{E}(t = 2 \tau) &=& \left \langle \langle \downarrow \vert \hat{U}^{\dag} \vert \downarrow \rangle
\langle \uparrow \vert \hat{U} \vert \uparrow \rangle \right \rangle_B \approx \\
\nonumber
&& \prod_{nm} 
\left \langle \langle \downarrow \vert \hat{U}_{n, m}^{\dag} \vert \downarrow \rangle
\langle \uparrow \vert \hat{U}_{n, m} \vert \uparrow \rangle \right \rangle_B,
\end{eqnarray}
where $v_{E}(t)$ is the spin echo decay.
We use heuristics to find the pairs with significant contributions and ignore the rest in order to make the calculation more efficient.

\begin{figure}
\includegraphics[width=3.5in]{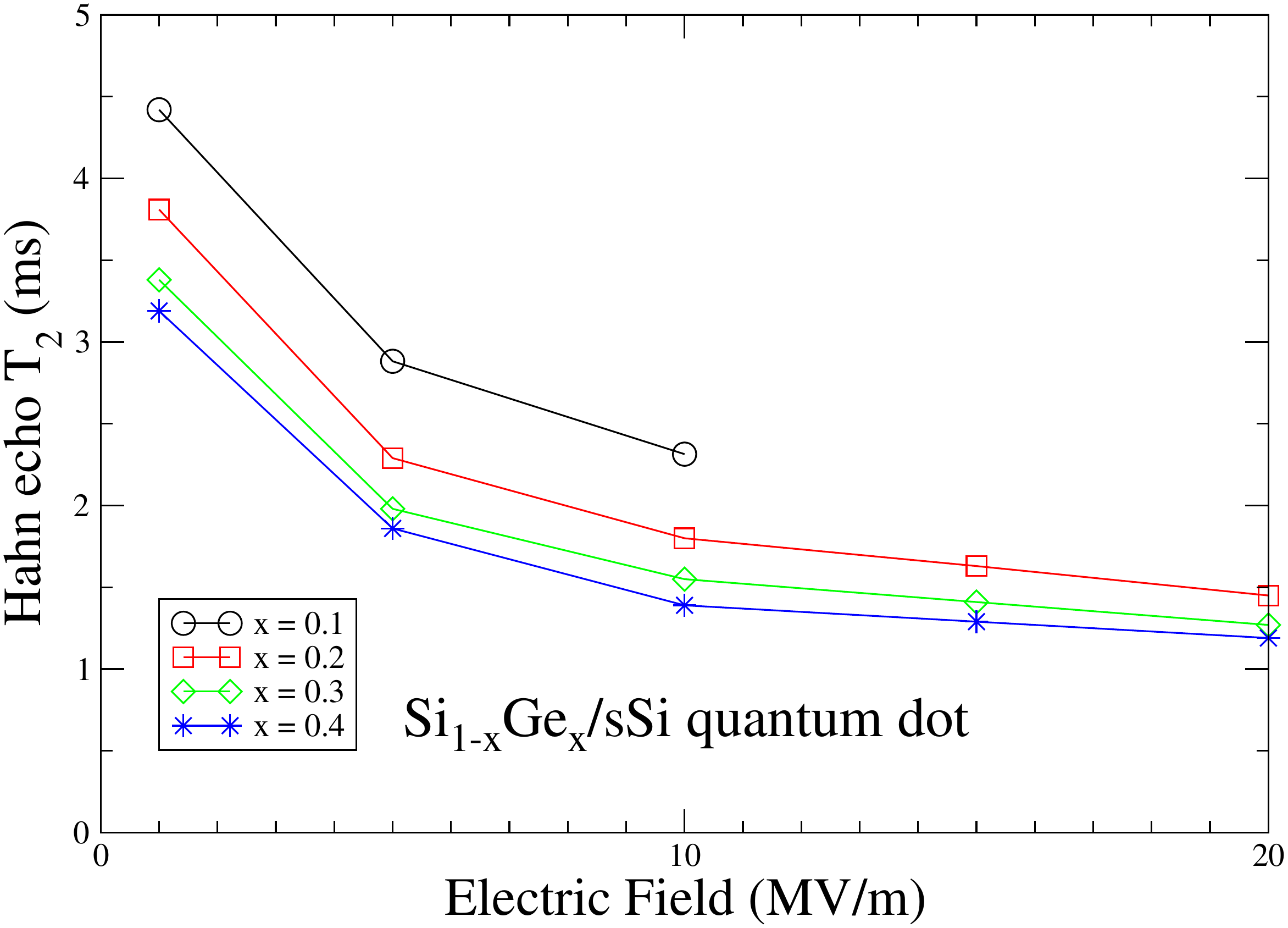}
\caption{
\label{NaturalSiGeT2s}
Hahn echo $T_2$ times for various Ge concentrations and electric
fields calculated for a SiGe/Si quantum dot electron
due to a naturally occurring abundance of $^{73}$Ge and no $^{29}$Si.
   The applied magnetic field is taken to be perpendicular to
the lateral direction.
}
\end{figure}

Our theoretical Hahn echo results for various SiGe/Si
quantum dot scenarios with decoherence induced by natural Ge are shown
in Fig.~\ref{NaturalSiGeT2s}.  We study the effects of different Ge
concentrations and different magnitudes of an applied electric field.
The electric field pulls the quantum dot deeper into the SiGe layer so
that $T_2$ times decrease.  An increase in concentration of the decoherence-inducing Ge causes $T_2$ to decrease but this is mitigated to some extent by the corresponding reduction in the overlap of the quantum dot wavefunction into the SiGe [Fig.~\ref{Fig:Wavefunction}~(b)].
The Hahn echo decay due
to natural Ge is approximately of the form $\exp{(-t^4)}$.
This behavior was theoretically predicted~\cite{WitzelCluster, YaoSiGe} in GaAs quantum dots
and experimentally confirmed~\cite{YacobyGaAsHahnEcho}.  It is
understood as the lowest order contribution in an expansion in time of the log of the echo.  The decay is produced
by a large number of small contributions from many different bath spin
pairs, each well-approximated by this lowest order in time.  This tends to occur in quantum dots that
envelope a large number of bath spins but not in donors where there is
a frozen core effect (nuclear spins near the center of a donor cannot
flip-flop because of the difference in their Knight
shifts)~\cite{WitzelCluster}.   Here, the bath spins are on the fringe
of the quantum dot, but the effect is the same.

\begin{figure}
\includegraphics[width=3.5in]{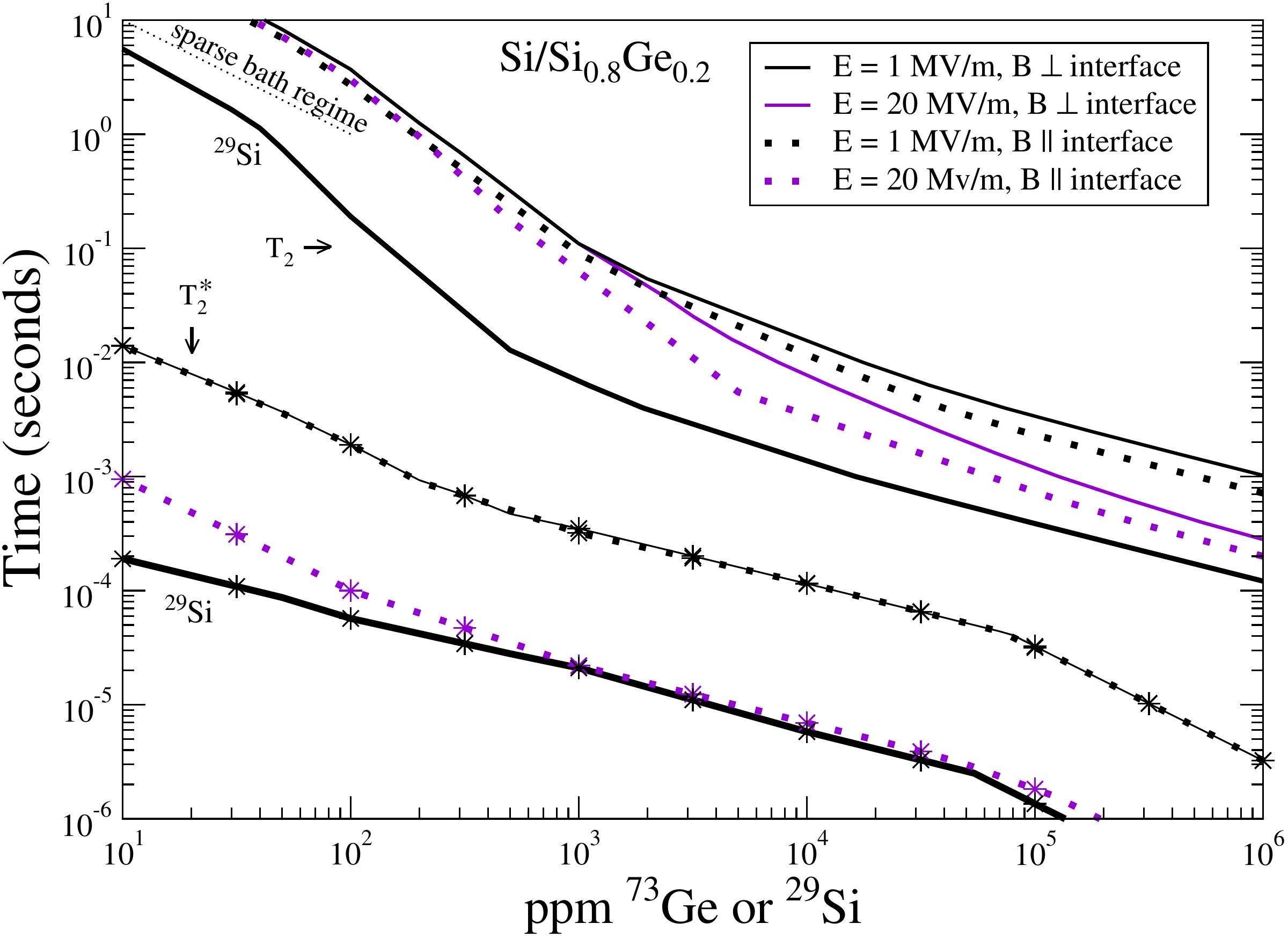}
\caption{
\label{T2vsF}
$T_2$ and $T_2^*$ for a SiGe/Si quantum dot electron
as a function of the fraction of the nonzero spin isotope of 
Ge ($^{73}$Ge) in the barrier layer assuming spinless Si.  Overlaid is the $T_2$ and $T_2^*$ of the quantum dot spin as a function of the fraction of the nonzero spin isotope of Si ($^{29}$Si) assuming spinless Ge.  The $^{29}$Si results are indicated with thick black curves and the rest are $^{73}$Ge results.  $T_2^*$
curves are indicated with star symbols.  The magnetic field direction
is perpendicular to the lateral direction for the solid curves and
parallel to the lateral direction for the dotted curves.  Colors
(color online) indicate different electric field strengths.
}
\end{figure}

$^{29}$Si enrichment has been used to attain long coherence times~\cite{Abe10, Tyryshkin50ppm}.  It is also possible, in principle, to simply remove the decoherence-inducing Ge spins via
isotopic enrichment.  
In Fig.~\ref{T2vsF}, we examine $T_2$ and $T_2^*$ times as a function
of various fractions of $^{73}$Ge or $^{29}$Si nuclei.  $T_2^*$ is
the $\exp{(-1)}$ decay time of a freely evolving ensemble of
similar quantum dots with different random spatial and state configurations of the spin bath (the inhomogeneous broadening effect).
For coherent control, gate operations should be fast relative to $T_2^*$.  $T_2$ is relevant for quantum memory.

For a given $^{29}$Si concentration it would be sensible to enrich the
$^{73}$Ge to attain a comparable $T_2$.
We observe that the effects of the electric field strength and
magnetic field direction upon $T_2$ are significant at higher
concentrations of $^{73}$Ge, but are not so relevant at low
concentrations.  In a sparse bath, such effects are minor.
The magnetic field direction has little effect upon
$T_2^*$.  At moderate to high concentrations isotropic contact hyperfine
interactions dominate $T_2^*$ and at low concentrations the system is
almost rotationally invariant.  The electric field, however, does
impact $T_2^*$ significantly since it has a strong impact upon the
contact hyperfine interactions.  At an electric field of
$20~\mbox{MV/m}$, the $^{73}$Ge $T_2^{*}$ nearly approaches that of
$^{29}$Si as a function of their respective isotopic fractions.  Although the overlap
into SiGe at this field and $x=0.2$ Ge concentration is only about
$10\%$, this is counter-balanced by the order of magnitude increase of
$\eta$ for the charge concentration at $^{73}$Ge nuclear sites.

The low concentration regime has a different character than the natural isotopic
concentration regime.  
In the low concentration limit, the geometric details of the problem become irrelevent and 
the effective Hamiltonian scales as $1/R^3$ (all dipolar interactions) so that time scales inversely with spin concentration.  This is denoted by the dashed line of Fig.~\ref{T2vsF}.
The generic problem in this regime is well-studied in Refs.~\onlinecite{WitzelDonors, WitzelDonorsLong}.
From these studies we note that there is substantial sample to sample variation such that the $T_2$ of a given qubit may differ substantially from the ensemble average $T_2$ that we report.

\section{Summary}
Decoherence of quantum dot electron spin in Si confined by SiGe barriers containing $^{73}$Ge are calculated.  The spatial extent of the wave function is calculated using a tight binding approach, NEMO3D, and the time dependence of the spin decoherence is calculated using a cluster expansion approach.  
The details of the SiGe penetration of the wavefunction are important for determining the isotropic contact hyperfind and anisotropic hyperfine interactions with the $^{73}$Ge.
Naturally abundant concentrations of $^{73}$Ge are predicted to limit decoherence times, $T_2$, to the order of milliseconds, many orders of magnitude below that experimentally demonstrated in enriched silicon ($^{28}$Si).  The decoherence time is predicted to increase rapidly through enrichment with zero nuclear spin Ge isotopes and reach decoherence times longer than that demonstrated in silicon (about $1$ second) at $^{73}$Ge concentrations below 200~ppm.  That is, other mechanisms in Si will likely limit the decoherence time at these enrichment levels.

\begin{acknowledgments}
Sandia National Laboratories is a
multiprogram laboratory operated by Sandia Corporation, a wholly
owned subsidiary of Lockheed Martin Corporation, for the U.S. Department
of Energy's National Nuclear Security Administration under contract
DE-AC04-94AL85000.
RR acknowledges Gerhard Klimeck for the NEMO 3D code.
WW acknowledges Jacob Taylor for noting the potential significance of 
quadrupole interactions that are briefly discussed in this paper.
\end{acknowledgments}

\end{document}